\documentclass[12pt]{article}

\usepackage{latexsym}

\begin{document}

\title{A class of homogeneous scalar-tensor cosmologies \\with a radiation 
fluid }

\author{
     Stoytcho S. Yazadjiev \thanks{E-mail: yazad@phys.uni-sofia.bg}\\
{\footnotesize  Department of Theoretical Physics,
                Faculty of Physics, Sofia University,}\\
{\footnotesize  5 James Bourchier Boulevard, Sofia~1164, Bulgaria }\\
}

\date{}

\maketitle

\begin{abstract}
We present a new class of exact homogeneous cosmological solutions
with a radiation fluid for all scalar-tensor theories. The
solutions belong to Bianchi type $VI_{h}$  cosmologies. Explicit
examples of nonsingular homogeneous scalar-tensor cosmologies are
also given.
\end{abstract}


\sloppy
\renewcommand{\baselinestretch}{1.3} %
\newcommand{\sla}[1]{{\hspace{1pt}/\!\!\!\hspace{-.5pt}#1\,\,\,}\!\!}
\newcommand{\db}{\,\,{\bar {}\!\!d}\!\,\hspace{0.5pt}}
\newcommand{\partb}{\,\,{\bar {}\!\!\!\partial}\!\,\hspace{0.5pt}}
\newcommand{\dsla}{\partb}
\newcommand{\eql}{e _{q \leftarrow x}}
\newcommand{\eqr}{e _{q \rightarrow x}}
\newcommand{\ite}{\int^{t}_{t_1}}
\newcommand{\itz}{\int^{t_2}_{t_1}}
\newcommand{\itd}{\int^{t_2}_{t}}
\newcommand{\lfrac}[2]{{#1}/{#2}}
\newcommand{\dV}{d^4V\!\!ol}
\newcommand{\ben}{\begin{eqnarray}}
\newcommand{\een}{\end{eqnarray}}
\newcommand{\la}{\label}


In  recent years, scalar-tensor theories attracted much attention
in many areas of gravitational physics and cosmology. From
theoretical point of view, it should be noted that these theories
arise naturally form the low energy limit of string theory
\cite{DP},\cite{GV}. In scalar-tensor theories, gravity is
mediated not only by the space-time metric but also by a scalar
field. The progress in the understanding of scalar-tensor
theories is closely connected with finding of exact solutions.
The aim of the present work is to present a new class of exact
homogeneous and anisotropic scalar-tensor cosmologies with a
radiation fluid. The found solutions belong to Bianchi type
$VI_{h}$ and, therefore, are different form previously known in
the literature homogeneous solutions which belong to either
Bianchi $I$, $V$, $IX$ or Kantowski-Sachs cosmologies
\cite{B}-\cite{CN}.

Scalar-tensor theories are described by the following action in
the Jordan frame \cite{W1}-\cite{EFDP}:
\begin{eqnarray} \label{JFA}
S = {1\over 16\pi G_{*}} \int d^4x \sqrt{-{\tilde
g}}\left({F(\Phi)\tilde R} - Z(\Phi){\tilde
g}^{\mu\nu}\partial_{\mu}\Phi
\partial_{\nu}\Phi  \right. \nonumber  \\ \left. -2 U(\Phi) \right) +
S_{m}\left[\Psi_{m};{\tilde g}_{\mu\nu}\right] .
\end{eqnarray}

Here, $G_{*}$ is the bare gravitational constant, ${\tilde R}$ is
the Ricci scalar curvature with respect to the space-time metric
${\tilde g}_{\mu\nu}$. The dynamics of the scalar field $\Phi$
depends on the functions $F(\Phi)$, $Z(\Phi)$ and $U(\Phi)$. In
order for the gravitons  to carry positive energy the function
$F(\Phi)$ must be positive. The nonnegativity of the energy of
the dilaton field requires that $2F(\Phi)Z(\Phi) +
3[dF(\Phi)/d\Phi]^2 \ge 0$. The action of matter depends on the
material fields $\Psi_{m}$ and the space-time metric ${\tilde
g}_{\mu\nu}$ but does not involve the scalar field $\Phi$ in
order for the weak equivalence principle to be satisfied. It
should be mentioned that the most used parametrization in the
literature is Brans-Dicke one, corresponding to $F(\Phi)=\Phi$
and $Z(\Phi)= \omega(\Phi)/\Phi$.

It is much clearer to analyze the equations in the so-called
Einstein frame. Let us introduce the new variables $g_{\mu\nu}$
and $\varphi$, and define

\begin{eqnarray}\label{GSD}
g_{\mu\nu} = F(\Phi){\tilde g}_{\mu\nu} \,\,\, , \nonumber  \\
 \left(d\varphi \over d\Phi \right)^2 = {3\over
4}\left({d\ln(F(\Phi))\over d\Phi } \right)^2 + {Z(\Phi)\over 2
F(\Phi)} \,\,\, , \\
 {\cal A}(\varphi) = F^{-1/2}(\Phi) \,\,\, ,\nonumber \\
2V(\varphi) = U(\Phi)F^{-2}(\Phi) \,\,\, .\nonumber
\end{eqnarray}

From now on we will refer to $g_{\mu\nu}$ and $\varphi$ as
Einstein frame metric and dilaton field, respectively.

In the Einstein frame the action (\ref{JFA}) takes the form

\begin{eqnarray}
S= {1\over 16\pi G_{*}}\int d^4x \sqrt{-g} \left(R -
2g^{\mu\nu}\partial_{\mu}\varphi \partial_{\nu}\varphi -
4V(\varphi)\right) \nonumber \\ + S_{m}[\Psi_{m}; {\cal
A}^{2}(\varphi)g_{\mu\nu}]
\end{eqnarray}

where $R$ is the Ricci scalar curvature with respect to the
Einstein frame metric $g_{\mu\nu}$.

The Einstein frame field equations then are

\begin{eqnarray} \label{EFFE}
R_{\mu\nu} - {1\over 2}g_{\mu\nu}R = 8\pi G_{*} T_{\mu\nu}
 + 2\partial_{\mu}\varphi \partial_{\nu}\varphi \nonumber \\  -
g_{\mu\nu}g^{\alpha\beta}\partial_{\alpha}\varphi
\partial_{\beta}\varphi -2V(\varphi)g_{\mu\nu}  \,\,\, ,\nonumber
\end{eqnarray}

\begin{eqnarray}
 \nabla^{\mu}\nabla_{\mu}\varphi = - 4\pi G_{*} \alpha (\varphi)T
+ {dV(\varphi)\over d\varphi} \,\,\, ,
\end{eqnarray}

\begin{eqnarray}
\nabla_{\mu}T^{\mu}_{\nu} = \alpha
(\varphi)T\partial_{\nu}\varphi \,\,\, . \nonumber
 \end{eqnarray}

Here $\alpha(\varphi)= {d\ln({\cal  A}(\varphi))/ d\varphi}$ and
the Einstein frame energy-momentum tensor $T_{\mu\nu}$  is
related to the Jordan frame one ${\tilde T}_{\mu\nu}$ via
$T_{\mu\nu}= {\cal A}^2(\varphi){\tilde T}_{\mu\nu}$. In the case
of a perfect fluid one has

\begin{eqnarray}\label{DPTEJF}
\rho ={\cal A}^4(\varphi){\tilde \rho}, \nonumber \\
p={\cal A}^4(\varphi){\tilde p}, \nonumber \\
u_{\mu}= {\cal A}^{-1}(\varphi){\tilde u}_{\mu}.
\end{eqnarray}

In the present work we consider the case $V(\varphi)=0$.

We were able to find the following class of exact solutions of
(\ref{EFFE}) for a radiation fluid with an equation of state
$\rho = 3p$ (${\tilde \rho} = 3 {\tilde p}$):

\begin{eqnarray}
ds^2 &=& \sinh^{4}(at)\left(-dt^2 + dx^2 \right) \nonumber \\ +
&e^{nax/\sqrt{3}}& \left(\sinh^{1+n}(at)e^{\sqrt{3}ax}dy^2 +
\sinh^{1-n}(at)e^{-\sqrt{3}ax}dz^2 \right), \nonumber \\
8\pi G_{*}p &=& {1\over 2}\left(1-{n^2\over 3 } \right){a^2 \over
\sinh^{4}(at)}, \\
\varphi(t)&=& \varphi_{0} + {\lambda\over a}\ln|\tanh(at/2)|,
\nonumber\\
u_{\mu} &=& - \sinh^2(at)\delta^{0}_{\mu} . \nonumber
\end{eqnarray}

Here, $n$  is given by $n^2 = 9 - 4{\lambda^2 \over a^2}$. The
solutions depend on three parameters $\varphi_{0}$, $a$ and
$\lambda$ where $\varphi_{0}$ is arbitrary, $a>0$ and  ${3\over
2}\le {\lambda^2\over a^2}\le {9\over 4}$. The range of the
coordinates is

\begin{equation}
0 < t <\infty, \,\,\, -\infty < x,y,z < \infty .
\end{equation}

The presented exact homogeneous cosmologies are solutions of the
Einstein frame equations (\ref{EFFE}) and, therefore, are
solutions to all scalar-tensor theories. The metric has three
dimensional nonabelian group  of isometries $G_{3}$ \cite{KSHMC}
acting on three-dimensional orbits and belonging to Bianchi type
$VI_{h}$ with $h=-n^2/9$. The corresponding Jordan frame
solutions can be simply obtained  by using transformations
(\ref{GSD}) and (\ref{DPTEJF}).

Let us consider some  properties of the solutions in the Einstein
frame. There is evidently a big-bang singularity at $t=0$ where
the curvature invariants, energy density, pressure and the dilaton
field  diverge.

The expansion always takes place along the $x$-axis whereas the
time evolution of the remaining axises depends on $n$. When $\mid
n\mid <1$, both axises expand. For $\mid n\mid =1$ one of the
axises expand while the other remains static. In the last case
when $1<\mid n\mid\le \sqrt{3}$ one of the axises expands whereas
the other is in contraction.

Three types of singular behaviour can be distinguished. For $\mid
n \mid <1$ we have anisotropic point like singularity - all three
spacial directions shrink as the initial time $t=0$ is
approached. For $\mid n\mid =1$ the singularity is of finite line
type. This type of singular behaviour describes the contraction
of two spacial directions whereas the third spacial direction
neither contracts or expands with time. The infinite line
singular behaviour occurs for $1<\mid n\mid\le \sqrt{3}$. In this
case, one of the spacial directions expands when the initial time
$t=0$ is approached. In fact, near the singularity the dilaton
dominates over the fluid and our model approaches a Bianchi type
$I$ model  with a minimally coupled scalar field. Near the
singularity the spacetime metric has a Kasner-like form

\begin{eqnarray}
g_{\mu\nu} \sim diag (-1, \eta^{2\gamma_{1}}, \eta^{2\gamma_{2}},
\eta^{2\gamma_{3}} )
\end{eqnarray}

where $\eta$ is the proper time and

\begin{eqnarray}
\gamma_{1} = {2\over 3}, \,\,\, \gamma_{2} = {1+n \over 6},
\,\,\, \gamma_{3} = {1-n \over 6} .
\end{eqnarray}

The Kasner exponents, as it should be expected,  satisfy the
Belinskii-Khalatnikov relations

\begin{eqnarray}
\gamma_{1} + \gamma_{2} + \gamma_{3} = 1, \\
\gamma_{1}^2 + \gamma_{2}^2 + \gamma_{3}^2 = 1 - {2\over
9}{\lambda^2 \over a^2}.
\end{eqnarray}

With regard to the late evolution, the dilaton field rapidly goes
to a constant value as $t$ increases - in other words general
relativistic behaviour is recovered at late times in both
Einstein and Jordan frame.

In general, the properties of the solutions in the Jordan frame
will depend on the particular scalar-tensor theory (i.e. on the
particular function ${\cal A}(\varphi)$). For some scalar-tensor
theories, Jordan frame solutions can exhibit a drastically
different behaviour from that in the Einstein frame. In order to
demonstrate that we shall consider the following scalar-tensor
theory described by the functions $F(\Phi)=\Phi$ and $Z(\Phi)=
\left(1-3\kappa^2 + 3\kappa^2\Phi
\right)/2\kappa^2\Phi\left(1-\Phi\right) $ which correspond to
${\cal A}(\varphi)=\cosh(\kappa\varphi)$ where $\kappa > 0$.

Let us consider the  Jordan frame solution with $n=0$ (i.e Bianchi
$VI_{0}$ type) which is given by:

\begin{eqnarray}\label{JFBSOS}
\Phi(t)&=& {4\tanh^{3\kappa}(at/2) \over \left[ 1 +
\tanh^{3\kappa}(at/2)\right]^2 } , \nonumber \\ d{\tilde s}^2 &=&
\Phi^{-1}(t) \sinh^{4}(at)\left(-dt^2 + dx^2 \right)  \nonumber
\\  &+& \Phi^{-1}(t)\sinh(at) \left(e^{\sqrt{3}ax}dy^2
+ e^{-\sqrt{3}ax}dz^2 \right), \\
8\pi G_{*}{\tilde p}&=& 8\pi G_{*} p  \Phi^2(t)= {8a^2\over
\sinh^{4}(at)} {\tanh^{6\kappa}(at/2) \over \left[ 1 +
\tanh^{3\kappa}(at/2)\right]^4 }, \nonumber \\
{\tilde u}^{\mu}&=&\Phi^{1/2}(t)u^{\mu} =
{2\tanh^{3\kappa/2}(at/2) \over \left[ 1 +
\tanh^{3\kappa}(at/2)\right] }\sinh^{-2}(at)\delta_{0}^{\mu}.
\nonumber
\end{eqnarray}

The curvature invariants  ${\tilde I_{1}} = {\tilde
C}_{\mu\nu\alpha\beta}{\tilde C}^{\mu\nu\alpha\beta}$, ${\tilde
I_{2}} = {\tilde R}_{\mu\nu}{\tilde R}^{\mu\nu}$  and ${\tilde
I_{3}}= {\tilde R}^2$ for the space-time metric (\ref{JFBSOS})
are:

\begin{eqnarray}
{\tilde I_{1}} = 3 a^4\left[{\Phi(t)\over \sinh^6(t)}\right]^2
\left[\left(1- \sinh^2(at)\right)^2 - 9\cosh^2(at)\sinh^2(at)
\right],
\end{eqnarray}

\begin{eqnarray}
{\tilde I_{2}} = 9\left[\Phi(t)\over 2\sinh^4(at)\right]^2
\left[\partial^2_{t}\ln(\Phi^{-1}(t)\sinh^2(at)) -
a\coth(at)\partial_{t}\ln(\Phi^{-1}(t)\sinh(at))\right]^2
\nonumber
\\ + \left[\Phi(t)\over 2\sinh^4(at)\right]^2
\left[ \partial^2_{t}\ln(\Phi^{-1}(t)\sinh^4(at)) \nonumber
\right. \\ \left.+
\partial_{t}\ln(\Phi^{-1}(t)\sinh^4(at))\partial_{t}\ln(\Phi^{-1}(t)\sinh(at))
 - 3a^2 \right]^2  \nonumber \\ +   2\left[\Phi(t)\over
2\sinh^4(at)\right]^2\left[
\partial^2_{t}\ln(\Phi^{-1}(t)\sinh(at)) \nonumber  +
\left(\partial_{t}\ln(\Phi^{-1}(t)\sinh(at))\right)^2\right]^2  ,
\nonumber
\end{eqnarray}

\begin{eqnarray}
{\tilde I_{3}}= \left[\Phi(t)\over 2\sinh^4(at)\right]^2 \left[
2\partial^2_{t}\ln(\Phi^{-1}(t)\sinh^4(at)) +
4\partial^2_{t}\ln(\Phi^{-1}(t)\sinh(at)) \nonumber\right. + \\
\left. 3\left(\partial_{t}\ln(\Phi^{-1}(t)\sinh(at))\right)^2 -
3a^2 \right]^2 . \nonumber
\end{eqnarray}

All curvature invariants, the energy density and  the pressure are
regular everywhere for $\kappa \ge 2$ as can be seen. Therefore
the solution has no big-bang singularity  or any other curvature
singularity.

Moreover, it can be shown that the solution under consideration
is also  geodesically  complete for $\kappa \ge 2$. Since the
space-time possesses group of isometries $G_{3}$, there are three
conserved quantities along the Killing fields which generate
$G_{3}$:

\begin{eqnarray}
\Phi^{-1}(t)\sinh(at)e^{\sqrt{3}ax}{dy\over d\tau}=L_{1} ,
\nonumber \\
\Phi^{-1}(t)\sinh(at)e^{-\sqrt{3}ax}{dz\over d\tau}=L_{2},
 \\
\Phi^{-1}(t)\sinh^4(at){dx\over d\tau}- {\sqrt{3}\over 2}aL_{1}y
+ {\sqrt{3}\over 2}aL_{2}z =L_{3} ,\nonumber
\end{eqnarray}

where $\tau$ is the affine parameter.

Taking into account these conserved quantities, the geodesic
equations can be reduced to

\begin{eqnarray}\label{GE}
{dt\over d\tau}&=& \pm \left[\Phi(t)\over \sinh^4(at)\right]^{1/2}
M^{1/2}(t,x,y,z), \nonumber \\
{dy\over d\tau}&=& L_{1}{\Phi(t)\over \sinh(at)}e^{-\sqrt{3}ax} , \\
{dz\over d\tau}&=& L_{2}{\Phi(t)\over \sinh(at)}e^{\sqrt{3}ax} ,\nonumber 
\\
{dx\over d\tau}&=& {\Phi(t)\over \sinh^4(at)}\left[L_{3} +
{\sqrt{3}\over 2}aL_{1}y - {\sqrt{3}\over 2}aL_{2}z \right],
\nonumber
\end{eqnarray}

where

\begin{eqnarray}
&M&(t,x,y,z)=\epsilon +  L^2_{1}{\Phi(t)\over
\sinh(at)}e^{-\sqrt{3}ax} + L^2_{2}{\Phi(t)\over
\sinh(at)}e^{\sqrt{3}ax} \nonumber \\&+&  {\Phi(t)\over
\sinh^4(at)}\left(L_{3} + {\sqrt{3}\over 2}aL_{1}y -
{\sqrt{3}\over 2}aL_{2}z \right)^2 ,
\end{eqnarray}

$\epsilon=0$  and $\epsilon=1$ for null and timelike geodesics,
respectively.

Let us first consider the geodesics with $\epsilon=1$  and
$L_{1}=L_{2}=L_{3}=0$.  From the explicit form of $\Phi(t)$ one
sees that $\Phi(t)/\sinh^4(at)$ is a bounded function and we
obtain

\begin{equation}
|{dt\over d\tau}|= \left[\Phi(t)\over \sinh^4(at)\right]^{1/2} <
const .
\end{equation}

Therefore, the only danger is if $t(\tau)$ could reach the value
$t=0$ for finite value of the affine parameter \cite{NOTE}.
However, this can not occur since the integral

\begin{equation}\label{int}
\int_{0}^{t}\left[\Phi(t)\over \sinh^4(at)\right]^{-1/2} dt
\end{equation}

is divergent for $\kappa \ge 2$. So, we conclude  that the
geodesics under consideration are  complete.

The same reasons show that the geodesics with $L_{3}\ne 0$,
$L_{1}=L_{2}=0$ and $\epsilon =0,1$ are complete,too.

Let us consider now the geodesics for which at least  one of the
constants $L_{1}$ or $L_{2}$ is different from zero
($\epsilon=0,1$ and $L_{3}$ arbitrary). In this case the
derivatives $dt/d\tau$, $dx/d\tau$, $dy/d\tau$ and  $dz/d\tau$
could become singular for finite values of the affine parameter
only for $x(\tau)$. More precisely, the derivatives can turn
singular for finite values of the affine parameter only if
$x(\tau)$ could grow unboundedly to $+\infty$ when $L_{2}\ne 0$
and to $-\infty$ when $L_{1}\ne 0$. However, when $L_{2}\ne 0$,
$x(\tau)$ cannot grow unboundedly to $+\infty$ for finite values
of $\tau$ since the derivative $dx/d\tau$ becomes negative for
large $x(\tau)$ (large $L_{2}z(\tau)$). In the same way, when
$L_{1}\ne 0$, $x(\tau)$ cannot grow unboundedly to $-\infty$ for
finite values of $\tau$ since the derivative $dx/d\tau$ becomes
positive for large $|x(\tau)|$ (large $L_{1}y(\tau)$). We
conclude thus that all derivatives are finite for finite values
of the affine parameter. The only danger, therefore, is if
$t(\tau)$ could reach the value $t=0$ for finite value of $\tau$.
This cannot occur since the integral (\ref{int}) is divergent and
$M(t(\tau), x(\tau),y(\tau),z(\tau))$ is finite for finite values
of $\tau$.

So, we proved that all causal geodesics are complete. This result
is not, of course, in  contradiction with the famous singularity
theorems. By computing the components of the Ricci tensor we have

\begin{eqnarray}
{\tilde R}_{00} = {3\over 2}\left[a\coth(at)\partial_{t}\ln\left(
\Phi^{-1}(t)\sinh(at) \right)  \right. \nonumber \\ \left.-
\partial^2_{t}\ln\left( \Phi^{-1}(t)\sinh^2(at) \right)  \right].
\end{eqnarray}

We may see that the sign of ${\tilde R}_{00}$ changes. The strong
energy condition is thus violated, and this explains why the
singularity theorems can be evaded.

Summarizing, we have presented a new class of homogeneous and
anisotropic cosmological solutions with radiation fluid for all
scalar-tensor theories. The found solutions are of Bianchi type
$VI_{h}$. The behaviour of the solutions in the Einstein frame
has been described. The behaviour of the solutions in the
physical Jordan frame depends on the particular scalar-tensor
theory and can be very different from that in the Einstein frame.
As an illustration of that, explicit examples of geodesically
complete homogeneous scalar-tensor cosmologies have been given.

\bigskip

\bigskip

\noindent{\Large\bf Acknowledgments}

\vskip 0.3cm

I  would like to thank P. Nikolov and V. Rizov for discussions.


\begin{thebibliography}{31}
\bibitem{NOTE} It is instructive to consider the following
example. Let us take $\kappa = 4/3$. Then the function
$\Phi(t)/\sinh^4(at)$ is bounded and $|dt/ d\tau| < const$.
However, the integral (\ref{int}) is convergent for $\kappa=4/3$
and the geodesic with $L_{1}=L_{2}=L_{3}=0$ and $\epsilon=1$
reaches $t=0$ for a finite value of the affine parameter and
cannot be continued. The space-time is thus singular as the
curvature invariants show for $\kappa=4/3$.
\end{thebibliography}
\end{document}